\documentclass[11pt,letterpaper]{article}
\usepackage{naaclhlt2016}
\usepackage{times}
\usepackage{latexsym}
\usepackage{graphicx}
\usepackage{caption}
\usepackage{subcaption}
\usepackage{epsfig}
\usepackage{amsmath}
\usepackage{amssymb}
\usepackage{color}
\usepackage{booktabs}
\usepackage{fixltx2e}
\usepackage{xcolor}
\usepackage{multirow}
\usepackage{tabularx}
\usepackage{flushend}
\usepackage{textcomp}
\usepackage{booktabs}
\usepackage{setspace}

\naaclfinalcopy

\newcommand{\squishlist}{
  \begin{list}{$\bullet$}
    { \setlength{\itemsep}{0pt}      \setlength{\parsep}{3pt}
      \setlength{\topsep}{3pt}       \setlength{\partopsep}{0pt}
      \setlength{\leftmargin}{1.5em} \setlength{\labelwidth}{1em}
      \setlength{\labelsep}{0.5em} } }
\newcommand{\squishlisttwo}{
  \begin{list}{$\bullet$}
    { \setlength{\itemsep}{0pt}    \setlength{\parsep}{0pt}
      \setlength{\topsep}{0pt}     \setlength{\partopsep}{0pt}
      \setlength{\leftmargin}{0.9em} \setlength{\labelwidth}{0.5em}
      \setlength{\labelsep}{0.5em} } }

 \newcommand{\squishend}{
     \end{list} 
 }

\title{Know2Look: Commonsense Knowledge for Visual Search}

\author{Sreyasi Nag Chowdhury \and Niket Tandon \and Gerhard Weikum\\
    Max Planck Institute for Informatics\\
    Saarbr\"ucken, Germany\\
  {\tt sreyasi, ntandon, weikum@mpi-inf.mpg.de}}

\date{}

\begin{document}

\maketitle

\begin{abstract}
With the rise in popularity of social media, images accompanied by contextual text form a huge section of the web. However, search and retrieval of documents are still largely dependent on solely textual cues. Although visual cues have started to gain focus, the imperfection in object/scene detection do not lead to significantly improved results. We hypothesize that the use of background commonsense knowledge on query terms can significantly aid in retrieval of documents with associated images. To this end we deploy three different modalities - text, visual cues, and commonsense knowledge pertaining to the query - as a recipe for efficient search and retrieval.
\end{abstract}


\section{Introduction}

{\bf Motivation:}
Image retrieval by querying visual contents has been
on the agenda of the database, information retrieval,
multimedia, and computer vision communities for decades~\cite{liu2007survey,datta2008image}. 
Search engines like Baidu, Bing or Google perform
reasonably well on this task, but crucially rely on
textual cues that accompany an image: tags, caption,
URL string, adjacent text etc.

\begin{figure}
    \centering
    \begin{subfigure}{0.5\textwidth}
        \centering
		\begin{tabular}{@{}p{3cm}p{4cm}@{}}\\
		\setlength{\tabcolsep}{12pt} 
		\vspace{-0.7cm}{\footnotesize Detected visual objects:	traffic light, car, person, bicycle, bus, car, grille, radiator grille} & 						\raisebox{-6em}{\includegraphics[width=3.8cm]{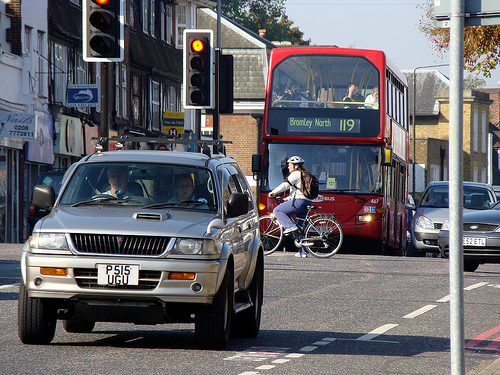}} \\
		\end{tabular}
	\caption{Good object detection}
    \label{goodObj}
    \end{subfigure}
    \begin{subfigure}{0.5\textwidth}
        \centering
		\begin{tabular}{@{}p{3cm}p{4cm}@{}}\\
		\setlength{\tabcolsep}{12pt} 
		\vspace{-0.3cm}{\footnotesize Detected visual objects: tv or monitor, cargo door, piano} & 
		\raisebox{-6em}{\includegraphics[width=3.8cm]{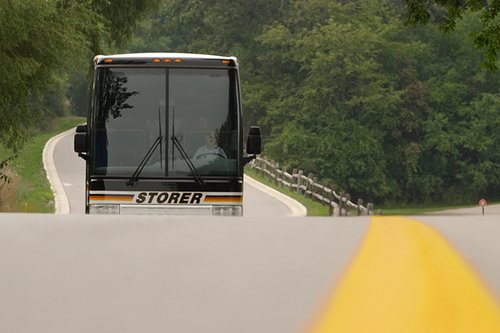}} \\
		\end{tabular}
		\caption{Poor object detection}
        \label{badObj}
    \end{subfigure}
    \vspace{0.2cm} 
\caption{Example cases where visual object detection may or may not aid in search and retrieval.}
\label{}
\vspace*{-0.5cm}
\end{figure}

In recent years, deep learning has led to a boost
in the quality of visual object recognition in images
with fine-grained object labels~\cite{simonyan2014very,lecun2015deep,mordvintsev2015inceptionism}.
Methods like LSDA~\cite{hoffman2014lsda} are trained on 
more than 15,000 classes of ImageNet~\cite{deng2009imagenet}
(which are mostly leaf-level synsets of WordNet~\cite{miller1995wordnet}),
and annotate newly seen images with class labels for 
bounding boxes of objects.
For the image in Figure~\ref{goodObj}, for example, object labels
{\em traffic light, car, person, bicycle} and {\em bus} have been recognized making it easily retrievable for queries with these concepts.
However, these labels come with uncertainty. 
For the image in Figure~\ref{badObj}, there is much higher
noise in its visual object labels; so querying by visual
labels would not work here.

\vspace{0.2cm}

\noindent
{\bf Opportunity and Challenge:}
These limitations of text-based search, on one hand, and 
visual-object search, on the other hand, suggest combining
the cues from text and vision for more effective retrieval.
Although each side of this combined feature space is 
incomplete and noisy, the hope is that the combination can
improve retrieval quality.

\begin{figure}[t]
\centering
\begin{tabular}{@{}p{3.6cm}p{5cm}@{}}
\\
\vspace{0.1cm} ``environment friendly traffic'' & \raisebox{-\totalheight}{\includegraphics[width=3.8cm]{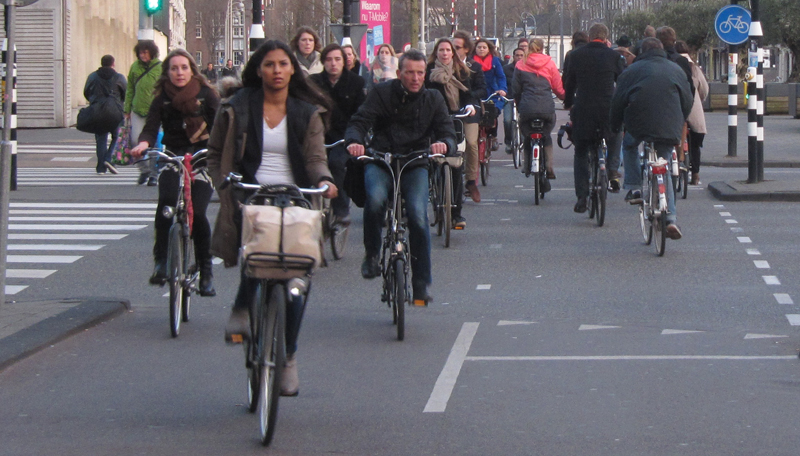}} \\
\vspace{0.1cm} ``downsides of mountaineering'' & 
\raisebox{-\totalheight}{\includegraphics[width=3.8cm]{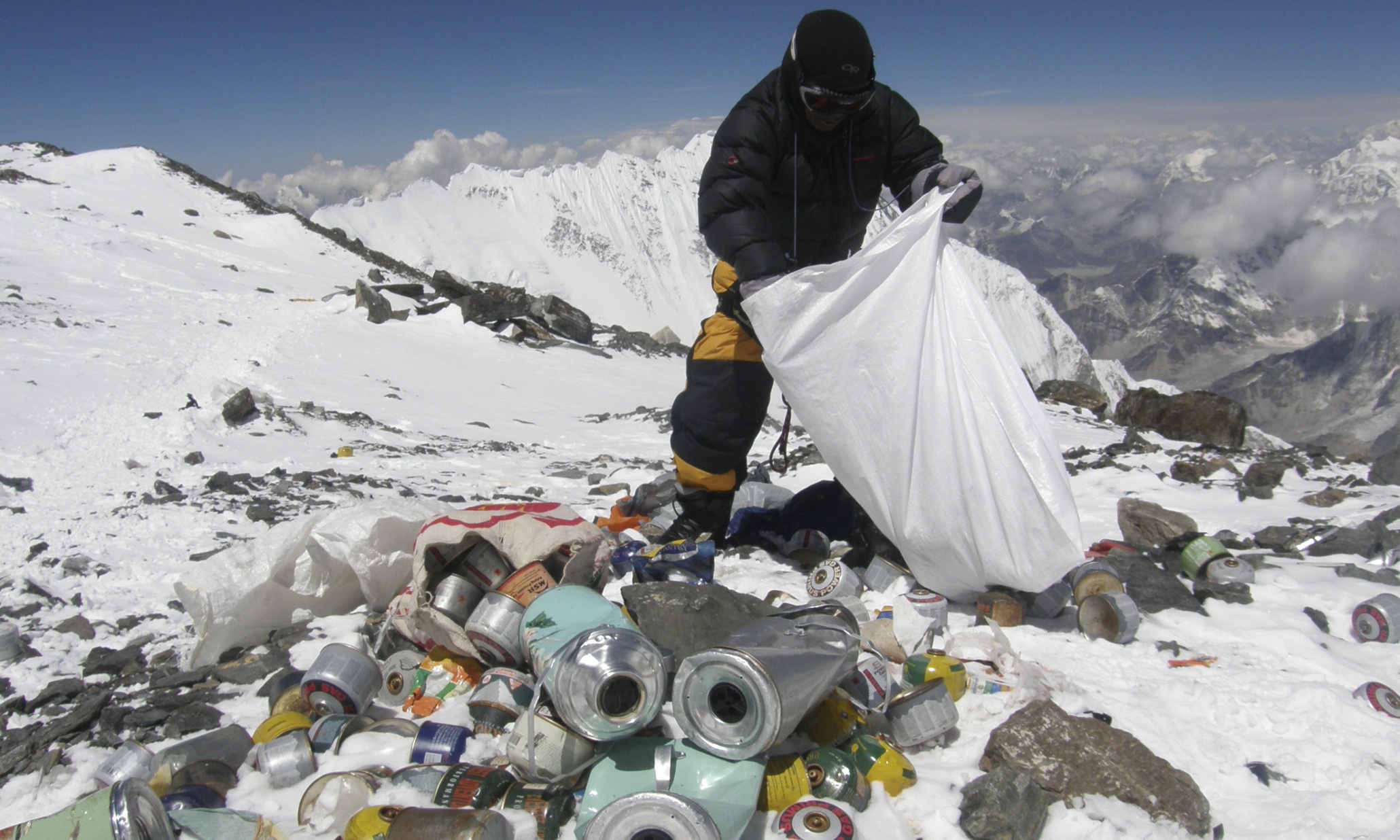}\hspace{0.5em}} \\
\vspace{0.1cm} ``street-side soulful music'' & 
\raisebox{-\totalheight}{\includegraphics[width=3.8cm]{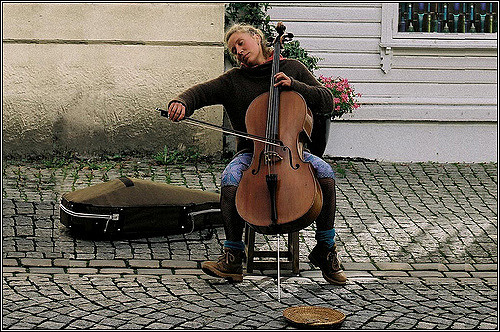}} \\
\end{tabular}
\vspace{0.2cm}
\caption{Sample queries containing abstract concepts and expected results of image retrieval.}
\label{sample}
\end{figure}

Unfortunately, images that show more sophisticated scenes,
or emotions evoked on the viewer are still out of reach.
Figure~\ref{sample} shows three examples, along with query formulations
that would likely consider these sample images as relevant results.
These answers would best be retrieved by queries with
abstract words (e.g. ``environment friendly'')
or activity words (e.g. ``traffic'') rather than words that
directly correspond to visual objects (e.g. ``car'' or ``bike'').
So there is a vocabulary gap, or even concept mismatch,
between what users want and express in queries and 
the visual and textual cues that come directly with an image.
This is the key problem addressed in this paper.


\vspace{0.2cm}

\noindent
{\bf Approach and Contribution:}
To bridge the concepts and vocabulary between
user queries and image features, we propose an
approach that harnesses commonsense knowledge (CSK).
Recent advances in automatic knowledge acquisition
have produced large collections of CSK:
physical (e.g. color or shape) as well as abstract (e.g.\ abilities)
properties of everyday objects (e.g.\ bike, bird, sofa, etc.)~\cite{tandon2014webchild}, subclass and part-whole relations between objects~\cite{tandon2016commonsense}, 
activities and their participants~\cite{tandon2015knowlywood}, and more.
This kind of knowledge allows us to establish relationships
between our example queries and observable objects or activities in the image.
For example, the following CSK triples establish relationships between {\em `backpack'}, {\em `tourist'} and {\em `travel map'}: {\small\tt(backpacks, are carried by, tourists)}, {\small\tt(tourists, use, travel maps)}. This allows for retrieval of images with generic queries like {\em ``travel with backpack''}.

This idea is worked out into a {\em query expansion model}
where we leverage a CSK knowledge base for
automatically generating additional query words.
Our model unifies three kinds of features:
{\em textual features} from the page context of an image,
{\em visual features} obtained from recognizing fine-grained object classes in
an image, 
and {\em CSK features} in the form of additional
properties of the concepts referred to by query words.
The weighing of the different features is crucial
for query-result ranking. To this end, we have devised a
method based on statistical language models \cite{zhai2008statistical}.

The paper's contribution can be characterized as follows.
We present the first model for incorporating CSK into image retrieval. 
We develop a full-fledged system architecture for this purpose, 
along with a query processor and an answer-ranking component. Our system 
{\em Know2Look}, uses commonsense {\em know}ledge to {\em look for} 
images relevant to a query by {\em looking at} the components of the 
images in greater detail. We further discuss experiments that compare our 
approach to state-of-the-art image search in various configurations.
Our approach substantially improves the query result quality.


\section{Related Work}

\noindent
{\bf Existing Commonsense Knowledge Bases:}
Traditionally commonsense knowledge bases were curated manually through experts~\cite{lenat1995cyc} or through crowd-sourcing~\cite{singh2002open}. Modern methods of CSK acquisition are automatic, either from test corpora~\cite{liu2004conceptnet} or from the web~\cite{tandon2014webchild}.

\vspace{0.2cm}

\noindent
{\bf Vision and NLP:}
Research at the intersection of Natural Language Processing and Computer Vision is in limelight in the recent past. There have been work on automatic image annotations~\cite{wang2014poodle}, description generation~\cite{vinyals2014show,Ordonez:2011:im2text,mitchell2012midge}, scene understanding~\cite{farhadi2010every}, image retrieval through natural language queries~\cite{malinowski2014multi} etc. 

\vspace{0.2cm}

\noindent
{\bf Commonsense knowledge from text and vision:}
There have been attempts for learning CSK from real images~\cite{chen2013neil} as well as from non-photo-realistic abstractions~\cite{vedantam2015learning}. Recent work have also leveraged CSK for visual verification of relational phrases~\cite{sadeghi2015viske} and for non-visual tasks like fill-in-the-blanks by intelligent agents~\cite{lin2015don}. Learning commonsense from visual cues continue to be a challenge in itself. The CSK used in our work is motivated by research on CSK acquisition from the web~\cite{tandon2014webchild}.


\section{Multimodal document retrieval}

Adjoining text of images may or may not explicitly annotate their visual contents. Search engines relying on only textual matches ignore information which may be solely available in the visual cues. 
Moreover, the intuition behind using CSK is that humans innately interpolate visual or textual information with associated latent knowledge for analysis and understanding. 
Hence we believe that leveraging CSK in addition to textual and visual information would take results closer to human users' preferences.  
In order to use such background knowledge, curating a CSK knowledge base is of primary importance. Since automatic acquisition of canonicalized CSK from the web can be costly, we conjecture that noisy subject-predicate-object (SPO) triples extracted through Open Information Extraction~\cite{banko2007open} may be used as CSK. We hypothesize that the combination of the noisy ingredients -- CSK, object-classes, and textual descriptions -- would create an ensemble effect providing for efficient search and retrieval.
We describe the components of our architecture in the following sections.

\subsection{Data, Knowledge and Features}

We consider a document $x$ from a collection $X$ with two kinds of features:
\squishlist
\item {\bf Visual features $xv_j$:}
labels of object classes recognized in the image, including their hypernyms
(e.g., king cobra, cobra, snake).
\item {\bf Textual features $xx_j$:}
words that occur in the text that accompanies the image, for example image caption.

\squishend
We assume that the two kinds of features can be combined into a single feature vector $x = \langle x_1 \dots x_M\rangle$ with hyper-parameters 
$\alpha_v$ and $\alpha_x$ to weigh visual vs. textual features.

CSK is denoted by a set $Y$ of triples $y_k (k=1..j)$
with components $ys_k, yp_k, yo_k$ ($s$ - subject, $p$ - predicate, $o$ - object).
Each component consists of one or more words.
This yields a feature vector $y_kj (j=1..M)$ for the triple $y_k$.

\subsection{Language Models for Ranking}

We study a variety of query-likelihood language models (LM) for ranking
documents $x$ with regard to a given query $q$.
We assume that a query is simply a set of keywords $q_i (i=1..L)$.
In the following we formulate equations for unigram LMs, which can be simply extended to bigram LMs by using word pairs instead of single ones.

\vspace*{0.2cm}
\noindent{}
{\bf Basic LM:}
\begin{equation}
    P_{basic}[q|x] = \prod_{i} P[q_i|x]
\end{equation}
where we set the weight of word $q_i$ in $x$ as follows:
\begin{equation}
    P[q_i|x] = \alpha_x P[q_i|xx_j]P[xx_j|x] + \alpha_v P[q_i|xv_j]P[xv_j|x]
\end{equation}
Here, $xx_j$ and $xv_j$ are unigrams in the textual or visual components of a document; $\alpha_x$ and $\alpha_v$ are hyper-parameters to weigh the textual and visual features respectively.

\vspace*{0.2cm}
\noindent{}
{\bf Smoothed LM:}
\begin{equation}
    P_{smoothed}[q|x] = \alpha P_{basic}[q|x] + (1-\alpha) P[q|B]
\end{equation}
where $B$ is a background corpus model and $P[q|B] = \prod_i P[q_i|B]$.
We use Flickr tags from the YFCC100M dataset~\cite{thomee2015yfcc100m} along with their frequency of occurrences as a background corpus.

\vspace*{0.2cm}
\noindent{}
{\bf Commonsense-aware LM} (a translation LM):
\begin{equation}
    P_{CS}[q|x] = \prod_{i}\bigg[ \dfrac{\sum_{k} P[q_i|y_k] P[y_k|x]}{|k|} \bigg]
\end{equation}
The summation ranges over all $y_k$ that can bridge the
query vocabulary with the image-feature vocabulary;
so both of the probabilities $P[q_i|y_k]$ and $P[y_k|x]$
must be non-zero. For example, when the query asks
for ``electric car'' and an image has features ``vehicle'' (visual)
and ``energy saving'' (textual), triples such as
{\small\tt (car, is a type of, vehicle)} and {\small\tt (electric engine, saves, energy)}
would have this property.
That is, we consider only commonsense triples that overlap with both
the query and the image features.


The probabilities $P[q_i|y_k]$ and $P[y_k|x]$ are estimated
based on the word-wise overlap between $q_i$ and $y_k$ and
$y_k$ and $x$, respectively. They also consider the 
confidence of the words in $y_k$ and $x$.


\vspace*{0.2cm}
\noindent
{\bf Mixture LM} (the final ranking LM):\\
Since a document $x$ can capture a query term
or its commonsense expansion, we formulate a mixture model for
the ranking of a document with respect to a query:
\begin{equation}
    P[q|x] = \beta_{CS} P_{CS}[q|x] + (1-\beta_{CS}) P_{smoothed}[q|x]
\end{equation}
where $\beta_{CS}$ is a hyper-parameter weighing the commonsense features of the expanded query.





\subsection{Feature Weights}

By casting all features into word-level 
unigrams,
we have a unified feature space with
hyper-parameters ($\alpha_x$, $\alpha_v$, and $\beta_{CS}$).
For this submission the hyper-parameters are manually chosen.


For weights of visual object class $xv_j$ of document $x$,
we consider the {\it confidence score} from LSDA~\cite{hoffman2014lsda}.
We extend these object classes with their hypernyms from WordNet which are set to the same confidence as their detected hyponyms.
Although not in common parlance this kind of expansion can also be considered as CSK.
We define the weight for a textual unigram $xx_j$ as its informativeness -- the inverse document frequency with respect to a background corpus (Flickr tags with frequencies).


The words in a CSK triple $y_k$ have non-uniform weights proportional to their similarity with the query words, their {\em idf} with respect to a background corpus, and the salience of their position -- boosting the weight of words in $s$ and $o$ components of $y$. The function computing similarity between two unigrams favors exact matches to partial matches.

%
%
%

\subsection{Example}
Query string: {\em travel with backpack}\\
Commonsense triples to expand query:

$t1$:{\small\tt(tourists, use, travel maps)}

$t2$:{\small\tt(tourists, carry, backpacks)}

$t3$:{\small\tt(backpack, is a type of, bag)}

\vspace{0.2cm}

\noindent
Say we have a document $x$ with features:

Textual -  ``A tourist reading a map by the road.''

Visual - person, bag, bottle, bus
%

\vspace{0.2cm}

\noindent
The query will now successfully retrieve the above document, whereas it would have been missed by text-only systems.
%
%
%
%
%
%
%



\section{Datasets}
For the purpose of demonstration we choose a topical domain -- {\em Tourism}. Our CSK knowledge base and image dataset obey this constraint.

\vspace{0.2cm}

\noindent
{\bf CSK acquisition through OpenIE:}
We consider a slice of Wikipedia pertaining to the domain {\em tourism} as the text corpus to extract CSK from. Nouns from the Wikipedia article titled `Tourism'(seed document) constitute our basic language model. We collect articles by traversing the Wiki Category hierarchy tree while pruning out those with substantial topic drift. 
The Jaccard Distance (Equation~\ref{jaccardDist}) of a document from the seed document is used as a metric for pruning.

\begin{small}
\begin{equation}
\label{jaccardDist}
Jaccard Distance = 1 - Weighted Jaccard Similarity
\end{equation}
\end{small}
where,
\begin{small}
\begin{equation}
\label{jaccard}
\begin{split}
Weighted Jaccard Similarity = \\
& \hspace{-1.5cm} \frac{\Sigma_n min[f(d_i,w_n), f(D, w_n)]}{\Sigma_n max[f(d_i,w_n), f(D, w_n)]}
\end{split}
\end{equation}
\end{small}

In Equation~\ref{jaccard}, acquired Wikipedia articles $d_i$ are compared to the seed document $D$; $f(d',w)$ is the frequency of occurrence of word $w$ in document $d'$.
For simplicity only articles with Jaccard distance of 1 from the seed document are pruned out. The corpus of domain-specific pages thus collected constitute \texttildelow 5000 Wikipedia articles.

The OpenIE tool ReVerb~\cite{fader2011identifying} run against our corpus
produces around 1 million noisy SPO triples. 
After filtering with our basic language model we have \texttildelow 22,000 moderately clean assertions.

\vspace{0.2cm}

\noindent
{\bf Image Dataset:}
For the purpose of experiments we construct our own image dataset.
\texttildelow 50,000 images with descriptions are collected from the following datasets: Flickr30k~\cite{young2014image}, Pascal Sentences~\cite{rashtchian2010collecting}, SBU Captioned Photo Dataset~\cite{Ordonez:2011:im2text}, and MSCOCO~\cite{lin2014microsoft}.
The images are collected by comparing their textual descriptions with our basic language model for {\em Tourism}. 
An existing object detection algorithm -- LSDA~\cite{hoffman2014lsda} -- is used for object detection in the images. 
The detected object classes are based on the 7000 leaf nodes of ImageNet~\cite{deng2009imagenet}. We also expand these classes by adding their super-classes or hypernyms with the same confidence score.

\vspace{0.2cm}

\noindent
{\bf Query Benchmark:}
We construct a benchmark of 20 queries from co-occurring Flickr tags from the YFCC100M dataset~\cite{thomee2015yfcc100m}. This benchmark is shown in Table~\ref{qBench}. Each query consists of two keywords that have appeared together with high frequency as user tags in Flickr images.

\begin{table}[]
\centering
\caption{Query Benchmark for evaluation}
\label{qBench}
\begin{tabular}{ll|ll}
\toprule
aircraft & international & diesel  & transport \\
airport  & vehicle       & dog     & park      \\
backpack & travel        & fish    & market    \\
ball     & park          & housing & town      \\
bench    & high          & lamp    & home      \\
bicycle  & road          & old     & clock     \\
bicycle  & trip          & road    & signal    \\
bird     & park          & table   & home      \\
boat     & tour          & tourist & bus       \\
bridge   & road          & van     & road     \\
\bottomrule
\end{tabular}
\end{table}

\section{Experiments}
{\bf Baseline}
Google search results on our image dataset form the baseline for the evaluation of {\em Know2Look}. We consider the results in two settings -- search only on original image caption (Vanilla Google), and on image captions along with detected object classes (Extended Google). The later is done 
to aid Google in its search by providing additional visual cues.
We exploit the domain restriction facility of Google search ({\em query string site:domain name}) to get Google search results explicitly on our dataset.

\vspace{0.2cm}

\noindent
{\bf Know2Look}
In addition to the setup for Extended Google, 
{\em Know2Look} also performs query expansion with CSK. In most cases we win over the baseline since CSK captures additional concepts related to query terms enhancing latent information that may be present in the images. 
We consider the top 10 retrieval results of the two baselines and {\em Know2Look} for the 20 queries in our query benchmark\footnote{http://mpi-inf.mpg.de/\texttildelow sreyasi/queries/evaluation.html}. We compare the three systems by Precision@10.
Table~\ref{tab:comparison} shows the values of Precision@10 averaged over 20 queries for each of the three systems -- {\em Know2Look} performs better than the baselines.

\begin{table}[]
\centering
\caption{Comparison of {\em Know2Look} with baselines}
\label{tab:comparison}
\setlength{\tabcolsep}{12pt}
\begin{tabular}{lllll}
\toprule
                & Average Precision@10  \\
\midrule
Vanilla Google  &   0.47        \\
Extended Google &   0.64        \\
Know2Look       &   0.85        \\
\bottomrule
\end{tabular}
\end{table}
\section{Conclusion}
\vspace{-0.2cm}
In this paper we propose the incorporation of commonsense knowledge for image retrieval. Our architecture, {\em Know2Look}, expands queries by related commonsense knowledge and retrieves images based on their visual and textual contents. By utilizing the visual and commonsense modalities we make search results more appealing to the humans than traditional text-only approaches. We support our claim by comparing {\em Know2Look} to Google search on our image data set. The proposed concept can be easily extrapolated to document retrieval. Moreover, in addition to using noisy OpenIE triples as commonsense knowledge, we aim to leverage existing commonsense knowledge bases for future evaluations of {\em Know2Look}.

\vfill

\noindent
{\bf Acknowledgment:}
We would like to thank Anna Rohrbach for her assistance with visual object detection of our image data set using LSDA. We also thank Ali Shah for his help with visualization of the evaluation results.

\clearpage



\onecolumn
\section*{Appendix}

The mathematical formulas, function definitions and details about hyper-parameters used are listed in tables \ref{tab:weights}, \ref{tab:params}, \ref{tab:LM}, and ~\ref{tab:functions}.

\vspace{1cm}

 \begin{table*}[h]
 \small
        \centering
        \begin{tabular}{@{}p{3.5cm}p{7.3cm}p{5.7cm}@{}}
        \toprule
                            & Formula & Description\\
        \midrule
        Textual feature weight    &  $P[xx_j|x] = \dfrac{idf(xx_j)}{\sum_\nu idf(xx_\nu)}$        &     The informativeness or weight of a word/phrase $xx_j$ in a document is captured by calculating it's idf in a large background corpus $\nu$.    \\
        \vspace{0.05cm}\\
        Visual feature weight     &  $P[xv_j|x] = \dfrac{conf(xv_j)}{\sum_\nu conf(xv_\nu)} \times
\dfrac{idf(xv_j)}{\sum_\nu idf(xv_\nu)}$        &   The weight of a object class $xv_j$ in a document is calculated by the product of it's confidence (from LSDA) and it's informativeness.       \\
        \vspace{0.05cm}\\
        CSK feature weight     &   $P[y_k|x]=\dfrac{\sum_i \sum_j sim(y_{kj},x_i)sal(y_{kj})inf(y_{kj})}{|i||j|}$        &    The relevance of a commonsense triple $y$ to a document is decided by the similarity of its words/phrases $y_k$ to the features of the document, the salience (or importance) of the match, and the informativeness of the word/phrase.    \\[0.5em]
        \bottomrule
        \end{tabular}
        \caption{Mathematical formulations of Feature Weights}
        \label{tab:weights}
    \end{table*}

    \vspace{1cm}
    

    \begin{table*}[h]
    \small
        \centering
        \begin{tabular}{@{}p{3cm}p{7cm}@{}}
        \toprule
        Hyper-parameter & Description\\
        \midrule
        $\alpha$        &   Weight of the basic document features; $(1-\alpha)$ being the weight for smoothing.        \\
        \vspace{0.05cm}\\
        $\alpha_x$      &  Weight associated with the textual features of a document.         \\
        \vspace{0.05cm}\\
        $\alpha_v$      &  Weight associated with the visual features of a document.         \\
        \vspace{0.05cm}\\
        $\beta_{CS}$      &     Weight pertaining to the commonsense knowledge features of an expanded document.      \\[0.5em]
        \bottomrule
        \end{tabular}
        \caption{Definition of Hyper-parameters}
        \label{tab:params}
    \end{table*}


\begin{table*}[]
\small
    \centering
        \begin{tabular}{@{}p{3.8cm}p{7cm}p{5.7cm}@{}}
        \toprule
                           & Formula & Description \\
        \midrule
        Basic LM             & $P_{basic}[q|x] = \prod_{i} P[q_i|x]$;        &   A unigram/bigram LM described by the probability of generation of a query $q$ from a document $x$. The weight of the $i^{th}$ word in $q$ is given by $P[q_i|x]$. The product over all words of the query ensures a conjunctive query. \\ 
                             &\raggedright $P[q_i|x] = \dfrac{\alpha_x}{|j|} \sum_j sim(q_i,xx_j)P[xx_j|x] + \text{\hspace{1.4cm}} \dfrac{\alpha_v}{|l|} \sum_l sim(q_i,xv_l)P[xv_l|x]$;      &    A word in the query may match with the textual or visual features of a document weighted by $\alpha_x$ and $\alpha_v$, and normalised with number of matches $|j|$ and $|l|$ respectively.\\
        \vspace{0.05cm}\\
        Smoothed LM          & $P_{smoothed}[q|x] = \alpha P_{basic}[q|x] + (1-\alpha) P[q|B]$; $P[q|B] = \prod_i P[q_i|B]$        &     The Basic LM after smoothing on background corpus $B$. The relative frequency of $q_i$ in $B$ ($P[q_i|B]$) is used for smoothing the LM.       \\
        \vspace{0.05cm}\\
        Commonsense-aware LM & $P_{CS}[q|x] = \prod_{i}\bigg[ \dfrac{\sum_{k} P[q_i|y_k] P[y_k|x]}{|k|} \bigg]$;        &   A translation LM describing the probability of generation of a query from the $k$ commonsense knowledge triples $y_k$. The summation over $k$ includes all triples bridging the gap between the query vocabulary and the document vocabulary; it is normalized by the total number of such triples.     \\
        & $P[q_i|y_k]=\sum_j sim(q_i,y_{kj})$ & The probability that the query word $q_i$ has been generated from the CSK triple $y_k$ is the sum of similarity scores between the two words/phrases, normalised by the number of words/phrases ($|j|$) in the CSK triples.
        
        \vspace{0.05cm}\\
        Mixture LM     & $P[q|x] = \beta_{CS} P_{CS}[q|x] + (1-\beta) P_{smoothed}[q|x]$        &    Combination of the weighted Commonsense-aware LM and Smoothed LM for ranking a document $x$ for a query $q$.      \\ [0.5em]
        \bottomrule
        \end{tabular}
        \caption{Mathematical formulations of Language Models for Ranking}
        \label{tab:LM}
    \end{table*}
    
    \vspace{0.3cm}


    \begin{table*}[]
    \small
        \centering
        \begin{tabular}{@{}p{2.5cm}p{7cm}p{5.7cm}@{}}
        \toprule
                            & Function & Description\\
        \midrule
        Confidence    &     $conf(w)$      &   A score output by the LSDA to depict the confidence of detection of an object class. The hypernyms of the detected visual object classes are assigned the same confidence score.        \\
        \vspace{0.05cm}\\
        Informative-ness     &    $inf(w)=idf_B(w)$       &   We measure informative-ness of a word by its idf value in a larger corpus, such that common terms are penalised.        \\
        \vspace{0.05cm}\\
        Similarity     &    $sim(w_1,w_2) = \dfrac{|substring(w_1,w_2)|}{max[length(w_1),length(w_2)]}$        &     This function calculates the amount of string overlap between $w_1$ and $w_2$.     \\
        \vspace{0.05cm}\\
        Salience     &  \begin{tabular}[t]{@{}l}
            $sal(w) = \lambda_s$ \hskip 1cm if $w \in subject$\\
            \hskip 1cm $= \lambda_p$ \hskip 1cm if $w \in predicate$\\
            \hskip 1cm $= \lambda_o$ \hskip 1cm if $w \in object$\\
            where $t_{csk} = \langle subject, predicate, object \rangle$ 
        \end{tabular}
        & The importance of the string match position in a commonsense knowledge triple $t_{csk}$ is captured by this function. Intuitively, the textual features in the subject and the object are more important that those in the predicate. Therefor we assign $\lambda_s = \lambda_o > \lambda_p$ and $\lambda_s+\lambda_p+\lambda_o=1$
        \\ [0.5em]
        \bottomrule
        \end{tabular}
        \caption{Function definitions}
        \label{tab:functions}
    \end{table*}

\end{document}